\documentstyle[12pt]{article}
\begin{document}
\newcommand{\be}{\begin{equation}}
\newcommand{\ee}{\end{equation}}
\newcommand{\bea}{\begin{eqnarray}}
\newcommand{\eea}{\end{eqnarray}}
\newcommand{\eps}{\varepsilon}
\newcommand{\Veff}%
{{\cal V}^{\mbox{\scriptsize p}}_{\mbox{\scriptsize eff}}}
\newcommand{\Afr}{A^{\mbox{\scriptsize fr}}}
\newcommand{\Bfr}{B^{\mbox{\scriptsize fr}}}
\newcommand{\inn}{\mbox{\scriptsize in}}
\newcommand{\exx}{\mbox{\scriptsize ex}}
\newcommand{\Fs}{\mbox{\scriptsize F}}
\vskip 5 true cm

\centerline{\bf Self-energy Effects in the Superfluidity of }
\centerline{\bf Neutron Matter }

\vskip 0.5 cm

\centerline{U.~Lombardo$^{1,2}$, P.~Schuck$^{3}$ and W.~Zuo$^{4}$}
\vskip 0.5 cm

\centerline{$^{1}$ Dipartimento di Fisica, 57 Corso Italia ,I-95129 Catania, 
Italy}
\centerline{$^{2}$ INFN-LNS, 44 Via S.~Sofia, I-95123 Catania, Italy}
\centerline{$^{3}$ Institut de Physique Nucl\'eaire, Universit\'e Paris-Sud,
F-91406 Orsay Cedex, France }
\centerline{$^{4}$ Institute of Modern Physics, Lanzhou, China} 
\vskip 0.5 cm

\centerline{Abstract}
\vskip 0.5 cm

The superfluidity of neutron matter in the channel $^1 S_0$ is studied by 
taking into account the effect of the ground-state correlations in the 
self-energy. To this purpose the gap equation has been solved within
the generalized Gorkov approach. A sizeable suppression of the energy gap 
is driven by the quasi-particle strength around the Fermi surface. 
\vskip 2.0 cm

\centerline{PACS Numbers: 26.60.+c, 74.20.Fg}

\newpage

              {\bf 1. Introduction}.
\vskip 0.3 cm
It is widely recognized that superfluidity is an extremely subtle process when
it is considered on an entirely microscopic level. Most of the calculations in
nuclear physics, at least concerning finite nuclei, are therefore based on 
phenomenological effective interactions. On the other hand a series of nuclear
matter and neutron matter calculations exists with the bare NN force used in
the gap equation \cite{LOM}. This is insofar the first step of a systematic
microscopic approach as it is well known that in the gap equation the bare
particle-particle interaction has to be taken to the lowest order and not a
$G$-matrix, i.e. a ladder summation, since the gap equation is in itself a
two-body equation \cite{MIG,RING}. The next terms in the effective pairing
interaction are screening terms. These are due to the possibility of medium
polarization which, when treated within the induced interaction theory 
\cite{BROW},
has turned out to strongly affect the pairing gap \cite{CLA,AINS,SCHU}. 
On the other hand
the self-energy corrections due to polarization effects have attracted much
less attention. Only very recently it has been stressed that the consideration 
of the quasi-particle strength can have a sizeable effect on the gap value
\cite{BAL}. Here we want to extend the latter work in investigating correlation
effects beyond the ones taken into account in Ref.~\cite{BAL} and we will see that
 they are indeed quite important. Our study must be considered as 
an intermediate step towards a fully consistent treatment where self-energy and
vertex corrections are taken into account on an absolutely equal footing.
The latter aspect may turn out to be decisive because sometimes strong 
cancellation between the two contributions can occur (see e.g. discussion
in Ref.~\cite{SCHK}). All those 
considerations are of great importance for a more microscopic understanding of
superfluidity in nuclei as well as in neutron stars, but also for a 
precise estimate of the pairing gap. In fact, pairing in exotic nuclei  at
present is studied with effective density dependent interactions modeled 
on pairing calculations in neutron matter \cite{SCHK,BERT}.  
In the latter systems
it is well known that superfluidity drives their rotational dynamics 
\cite{ALP} as well as their cooling \cite{RIPE}. In this work we shall study pure neutron matter but we
expect that very analogous effects will occur in symmetric nuclear matter.  
\vskip 0.3 cm
              {\bf 2. Neutron self-energy}.
\vskip 0.3 cm
Let us first discuss some properties of the single-particle self-energy 
$\Sigma_p(\omega)$ of neutron matter. In the Brueckner approach \cite{MAH} 
the perturbative expansion of $\Sigma$ can be recast according to the number 
of hole lines as follows  
\be
\Sigma_p(\omega)  =  \Sigma^{1}_p(\omega)  +  \Sigma^{2}_p(\omega) + \dots 
\ee

The on-shell values of $\Sigma^1$ represent the Brueckner-Hartree-Fock (BHF) 
mean field and the on-shell values of $\Sigma^2$ give the so called 
rearrangement term, which gives largest contribution to the ground-state 
correlations. The off-shell values enter several physical properties of neutron
matter, including pairing. 
In terms of the self-energy one may calculate, at a given order of the hole-line
expansion, the quasi-particle energy, as solution $\omega_p$ of the 
equation
\be
\omega_p \,=\, \frac{p^2}{2m} - \epsilon_F + \Sigma_p(\omega_p)
\label{eq:qpe}
\ee
where $\epsilon_F$ is the Fermi energy. We  here neglect the imaginary part
of the self-energy.
The quasi-particle energy around the Fermi surface is obtained expanding the 
self-energy around $p=p_F$ and $\omega=0$: 
\be
\omega \,=\, \frac{p^2-p_F^2}{2m^*} \,=\, \frac{p^2-p_F^2}{2m} 
\frac{m}{m_e m_p}
\ee 
where $m^*$,$m_e$ and $m_p$ are the effective mass, $e$-mass and $p$-mass,
respectively. The latter two masses are defined as  
\be
m_e \,=\,  m \bigg[ 1 - \bigg( \frac{\partial \Sigma}{\partial\omega}\bigg)_F 
\bigg]
\label{eq:emas}
\ee
\be
m_p \,=\, m \bigg[ 1 + \frac{m}{p_F} \bigg( \frac{\partial \Sigma}{\partial p}
\bigg)_F \bigg]^{-1} 
\label{eq:kmas}
\ee 
The partial derivatives are evaluated at the Fermi surface.
The $k$-mass is related to the non-locality of the mean field and, 
if the self-energy is $\omega$-independent (static limit), it coincides 
with the effective mass. This quantity 
is of great interest whenever the momentum dependence of the mean field can 
give some effects such as transverse flows in heavy-ion collisions \cite{SAND}. 
The $e$-mass is related to the quasi-particle strength. This latter gives the 
discontinuity of the momentum distribution at the Fermi surface, and
measures the amount of correlations included in the considered approximation.

The definition of $m^*$, $k$-mass and $e$-mass can be extended to $p \neq p_F$
replacing in Eqs.~(\ref{eq:emas}) and (\ref{eq:kmas}) $p_F$ by $p$ and
$\omega=0$ by $\omega_p$. Their properties have been extensively studied in 
Ref.~\cite{MAH}.

The self-energy has been calculated in two approximations: up to first order 
(BHF) and up to the second order (EBHF) of the hole-line 
expansion in the framework of the Brueckner theory adopting the continuous 
choice \cite{ZUO}. The Argonne $V_{14}$ potential has been 
used for the bare interaction \cite{WIR}. The calculations have been performed 
for a range of Fermi momenta where the energy gap is expected to be the largest, 
i.e. $0.5 \le k_F \le 1.3 fm^{-1}$, corresponding to a density range 
from 0.0042 through 0.074 $fm^{-3}$. 
A typical result for the off-shell neutron self-energy $\Sigma_p(\omega)$
is plotted in Fig.~1. The contribution from the rearrangement term shows a
pronounced enhancement in the vicinity of the Fermi energy, which is to be
traced back to the high probability amplitude for particle-hole excitations 
near $\epsilon_F$ \cite{ZUO}. At high momenta this contribution vanishes. 
From $\Sigma_p(\omega)$ the effective masses  
are extracted according to Eq.s~(4) and (5). They are depicted in Fig.~2, where
the full calculation is compared to that including only the BHF self-energy.
We may distinguish two momentum intervals: at $k\approx k_F$ the momentum
dependence of the effective mass $m^*$ is characterized by a bump, whose peak 
value exceeds the value of the bare mass; far above $k_F$ the bare mass limit 
is approached. One should take into account that in this range of $k_F$ the 
neutron
density is quite small (at the maximum $k_F=1.3 fm^{-1} \rho=0.074 fm^{-3}$). 
This behaviour of the effective mass $m^*$ is due mostly to the e-mass, as shown
in the lower panel of Fig.~2. In both panels of Fig.~2 it is also reported, 
for comparison, the effective mass in the BHF limit (only $\Sigma^{1}$ 
included), 
which exhibits a much less pronounced bump at the Fermi energy. It is precisely 
the influence of this increased bump structure, which we want to investigate
here, since in Ref.~\cite{BAL} the rearrangement term has not been taken into
account.
\vskip 0.3 cm
 
              {\bf 3. Generalized gap equation }
\vskip 0.3 cm
The generalized BCS theory can be found in various textbooks on the 
many-body problem \cite{MIG,NOZ,ABRI}. Here we follow closely the formalism 
developed in \cite{MIG}, where the gap equation is written as
\be
\Delta_{p}(\omega) \,=\, - \int \frac{d^3 p'}{(2\pi)^3} \int \frac{d\omega}
{2\pi i} {\cal V}_{p,p'}(\omega,\omega') \Gamma_{p'}(\omega')\Delta_{p'}
(\omega')
\label{gge}
\ee
The kernel $\Gamma$ is defined as
\be
\Gamma_{p}(\omega) \,=\, G_{p}(-\omega)G_{p}^{s}(\omega) \,=\, 
 [G_{p}^{-1}(\omega) G_{p}^{-1}(-\omega) + \Delta_p^2 (\omega)]^{-1}  
\label{gam} 
\ee
The functions $G_{p}(\omega)$ and $G_{p}^{s}(\omega)$ are the nucleon 
propagators of neutron matter in the normal state and in the superfluid state, 
respectively . The $\omega$-symmetry in the two propagators is
to be traced back to the time-reversal invariance of the Cooper pairs.
The effective interaction ${\cal V}$ is the block of all irreducible diagrams 
of the interaction. The short-range correlations (ladder diagrams) are already taken into account by the gap
equation and do not appear in the irreducible block ${\cal V}$. 
The long-range components
have been studied in the context of the induced interaction approach 
\cite{AINS,SCHU}.
In the present note we only consider the first term in the perturbative 
expansion of ${\cal V}$,
namely the bare interaction, because we want to disentangle the influence of
correlations coming only from the self-energy expansion from the ones due to
the induced interaction. The complete 
solution of the generalized gap equation requires a further effort.
         
Assuming the pairing interaction to be identified with the bare interaction
$V_{p,p'}$, the energy gap does not depend on the energy (static limit), i.e. 
$\Delta_p(\omega)\equiv \Delta_p$. In this limit the self-energy corrections
are not expected to modify the analytical structure of the kernel 
$\Gamma_p(\omega)$ which now is an even function of energy: at each momentum 
$p$ there exist two symmetric poles $\pm \Omega_p$ in the complex $\omega$-plane. 
The $\omega$ integration can be performed as follows:
\be
\int \frac{d\omega}{2\pi i} \Gamma(\omega^2) \,=\, - \frac{{\cal Z}_p^2}{2\Omega
_p}
\ee
where we denote by ${\cal Z}_p^2$ the residue of the kernel at the pole 
$\Omega_p$. Since
the largest contribution to the integral is coming from the pole part of the
two Green's function, we expand the single particle propagator 
to first order in $\omega_p$ (see Eq.~(\ref{eq:qpe})), 
$G_p(\omega)\approx {\cal Z}_p \cdot (\omega -
\omega_p)^{-1}$, and therefore the denominator in Eq.~(\ref{gam}) becomes
\be
\Gamma^{-1}(\omega^2) \approx {\cal Z}_p^{-2} (\omega^2-\omega_p^2) + \Delta^2_p , 
\ee
where ${\cal Z}_p^{-2}$ is given by
\be
 {\cal Z}_p^{-2} \approx \bigg(\frac{\partial G^{-1}(\omega)}{\partial\omega}
\bigg)\bigg|_{\omega=\omega_p} \cdot
\bigg(\frac{\partial G^{-1}(-\omega)}{\partial\omega}\bigg)\bigg|_
{\omega=-\omega_p} = -\bigg[ 1-\frac{\partial\Sigma_p(\omega)}{\partial\omega}
\bigg]^2\bigg|_{\omega=\omega_p}  
\ee
and 
\be
\Omega_p \approx \sqrt{\omega_p^2 + {\cal Z}_p^2 \Delta_p^2}
\ee
In this approximation the generalized gap equation, Eq.~(6), becomes
\be
\tilde{\Delta}_{p} \,=\, - \frac{1}{2} \int \frac{d^3 p'}{(2\pi)^3} \frac{{\cal Z}_p V_{pp'}
{\cal Z}_{p'}}{\sqrt{\omega_{p'}^2 + \tilde{\Delta}_{p'}^2}} 
\,\,\,\, \tilde{\Delta}_{p'} 
\label{eq:AGE}
\ee
where $\tilde{\Delta}_p = {\cal Z}_p \Delta_p$ is the real pairing correction
to the quasi-particle energy spectrum.
The main difference from the BCS limit is the presence of the quasi-particle
strength, which is less than one in a small region around the Fermi surface
as we saw in the previous section. The pairing interaction turns out to be 
reduced in that region, where the Cooper pairs are mainly formed. This is the
way self-energy corrections come into play suppressing the pairing gap.  
As to the self-energy effects, Eq.~(\ref{eq:AGE}) is quite general because there
is no expansion in the self-energy so far. The self-energy corrections can be
taken at any order of approximation.

One may further restrict the $\omega$-integration to only the pole part at 
the Fermi energy, i.e. expanding the self-energy near the Fermi surface 
according to Eq.~(3). In this case the integration can  also be 
performed analytically and one easily obtains
\be
\tilde{\Delta}_p \,=\, - {\cal Z}_F^2 \int \frac{d^3 p'}{(2\pi)^3} 
\frac{V_{p,p'} \tilde{\Delta}_{p'}}{2\sqrt{p_F^2(p'-p_F)^2
/{m^*}^2 + \tilde{\Delta}_{p'}^2}}.
\ee
where ${\cal Z}_F$ is the quasi-particle strength at the Fermi surface and
coincides with the inverse of the $e$-mass defined by Eq.~(\ref{eq:emas}). It amounts to
the discontinuity of the momentum distribution at the Fermi surface and 
measures to content of correlations included in the model.
As is well known the pairing modifies the chemical potential which is 
calculated self-consistently with the gap equation from the closure equation
for the density of neutrons. In our approximation it is given by
\be 
\rho \,=\, 2 \int \frac{d^3 p}{(2\pi)^3} \int \frac{d\omega}{2\pi i} G_p^{s}(\omega^+)
\ee
\be
\approx 2 \int \frac{d^3 p}{(2\pi)^3} \frac{{\cal Z}_p}{2} 
(1-\frac{\omega_p}{\sqrt{\omega_p^2 + \tilde{\Delta}_{p}^2}})
\ee
The prefactor 2 is due to spin degeneracy.
Our numerical investigation is based on the solution of the two coupled gap 
equations, Eqs.(12) and (15), along with the self-energy approximated
to the second order of the hole-line expansion.  
\vskip 0.3 cm
{\bf 2. Results and conclusions}.
\vskip 0.3 cm
The Argonne $A14$ potential has been adopted as pairing interaction which is consistent 
with the self-energy data where the same force has been used.  
The gap equation has been solved in the form of Eq.~(12), coupled with
Eq.~(15). This is a quite satisfactory approximation, especially in view of
studying the self-energy effects.    
  The results are reported in Fig.~3 for a set of different
  $k_F$-values. The solid line represents the solution 
  of the gap equation in the standard BCS limit with the free single-particle spectrum. 
  This is very close to the prediction obtained replacing the bare mass by
  the effective mass calculated in the BHF but still keeping ${\cal Z}=1$
  (patterned line). This similarity stems from the fact that at the Fermi 
  surface $m^*/m$ from BHF is close to one as shown in Fig.~2. 
  The self-energy effects are estimated in two approximations. In the first 
  one $m^*$ and the ${\cal Z}$-factor are calculated from the approximation 
  $\Sigma=\Sigma^{1}$ in a BHF code. In the 
  considered density domain the {\cal Z}-factor is around $\approx 0.9$.  
  Despite its moderate reduction a strong suppression of the gap is 
  obtained as shown in Fig.~3 (upper long-dashed line).
  It is due to the exponential dependence of the gap on all quantities. 
  Still a further but more moderate reduction is
  obtained when the rearrangement term is included in the second 
  approximation $\Sigma\approx\Sigma^{1} +\Sigma^{2} $ (short-dashed line). 
  The smaller ${\cal Z}$-factor (${\cal Z} \approx .83 $ at  $k_F=.8 fm^{-1}$) 
  is to a certain extent
  counterbalanced by an increase of the effective mass ($ m^* /m \approx
  1.2 $ at the same $k_F$).

  Self-energy corrections are mostly concentrated around the Fermi surface; 
  therefore it is not appropriate to use Eq.~(13) for a quantitative 
  prediction of the gap, since it extends their effect beyond the Fermi
  surface. The self-energy effect turns out to be overestimated by Eq.~(13)
  as we checked numerically. In Fig.~3 the results are reported for the
  two adopted approximations of the self-energy (lower long-dashed and short-dashed
  lines, respectively). 
      
  In Ref.~\cite{BAL} the selfenergy effect has also been investigated
  within the generalized gap equation but the selfenergy has been
  considered only at the level of BHF approximation. However the approximations
  adopted in Ref.~\cite{BAL} for solving the gap equation understimate the  
  correlation effects so that a moderate reduction is obtained.  
  In the present approach the reduction is more pronounced with only 
  $\Sigma^{1}$. The inclusion of  
  the rearrangement term $\Sigma^{2}$ brings about a further non negligible
  reduction.  

  In conclusion we have shown that the superfluidity of a strongly correlated
  Fermi system requires to be described in the context of the generalized Gorkov
  approach. The fact that the quasiparticle strength can be significantly
  smaller than one cannot be counterbalanced by a corresponding enhancement
  of the effective mass. Moreover we have shown that reliable predictions
  from the generalize gap equation can only be obtained if the correlation
  effects are fully taken into account. We have treated here neutron matter
  but we expect that the selfenergy effects on the gap are very similar in
  symmetric nuclear matter. A next important step forward will be to include
  not only selfenergy effects but also, on an equal footing, vertex
  corrections. This shall be studied in a future work.

  Acknowledgments

  Discussions with Drs. M. Baldo and H.-J. Schulze are gratefully acknowledged. 

  This work has been supported in part by {\it Regroupement de Recherche, 
  Noyaux Exotiques}, CNRS-IN2P3, France.
                                 

\newpage
\vskip 2cm
\centerline{Figure captions}
\begin{itemize}
\item{Fig.1} One hole-line off-shell $\Sigma_k^1(\omega)$ (solid line) and 
             two-hole line
             off-shell $\Sigma_k^2(\omega)$ (dashed line) at a fixed momentum
             $k$ for three neutron densities $\rho = k_F^3/3\pi^2$.\\ 
\item{Fig.2} Effective mass (upper panels) and $e$-mass (lower panels) for 
             three densities $\rho = k_F^3/3\pi^2$ in the BHF approximation
             (dashed line) and EBHF approximation (solid line).\\
\item{Fig.3} Energy gap in different approximations 
             ($\tilde \Delta$ in the text). Solid line: free 
             single-particle spectrum; patterned line: effective mass from
             BHF approximation and ${\cal Z}=1$; upper (lower) long-dashed line:
             solving the gap equation in the form of Eq.~(12) (Eq.~(13)) with
             BHF effective mass and ${\cal Z}$; upper (lower) short-dashed line:
             solving the gap equation in the form of Eq.~(12) (Eq.~(13)) with
             EBHF effective mass and ${\cal Z}$.  
\end{itemize}
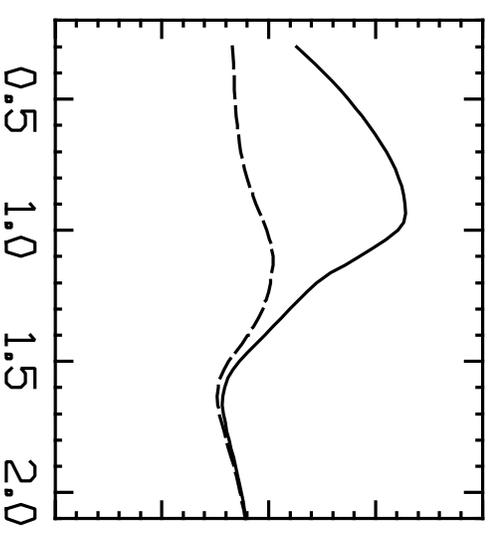
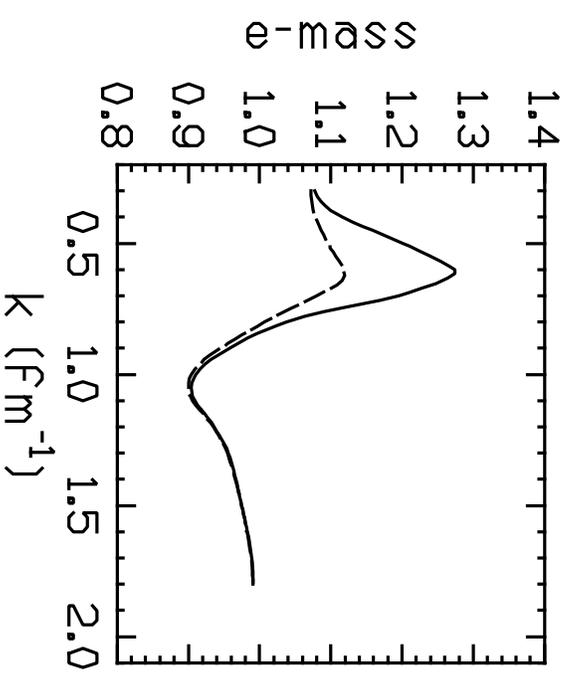
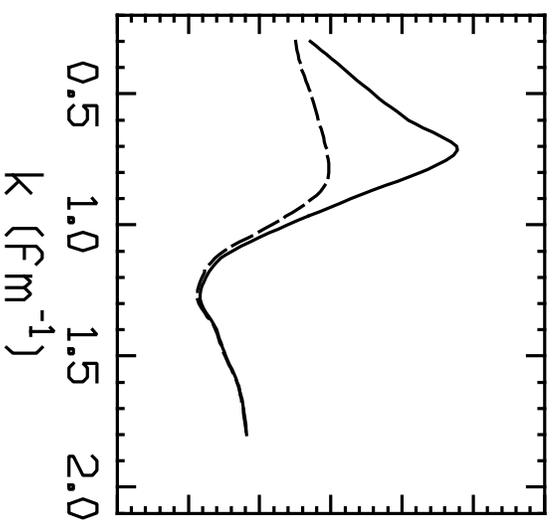
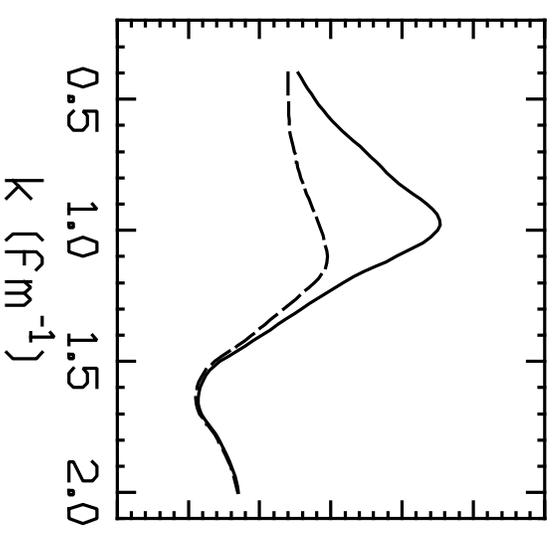
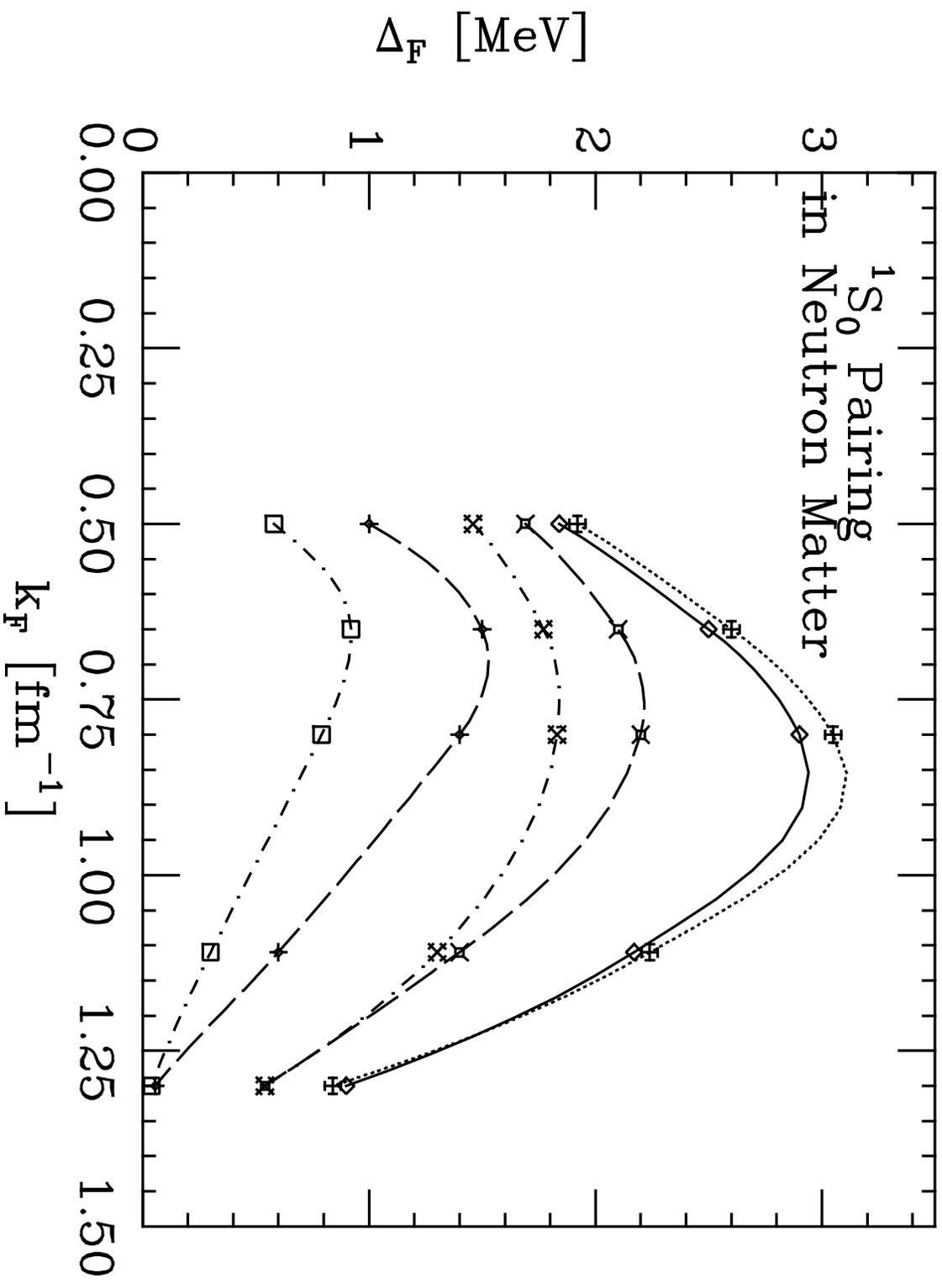
\end{document}